\documentclass[twocolumn,showpacs,amssymb,prd]{revtex4}

\usepackage{graphicx}% Include figure files
\usepackage{dcolumn}% Align table columns on decimal point
\usepackage{bm}% bold math
\usepackage{epsfig}
\begin{document}
\newcommand{\gsim}{\hbox{\rlap{$^>$}$_\sim$}}
\newcommand{\lsim}{\hbox{\rlap{$^<$}$_\sim$}}

\title{The superluminal motion of the jet launched in GW170817,\\
     the Hubble constant, and critical tests of gamma ray bursts theory}

\author{Shlomo Dado and Arnon Dar}
\affiliation{Physics Department, Technion, Haifa, Israel} 

\begin{abstract}
The direction of the axis of the orbital motion of the merging binary 
neutron stars in the GW170817 event coincided with that of the apparent 
superluminal jet, which produced the short hard gamma ray burst (SHB) 
170817A. It supports the local value of the Hubble constant provided by 
standard candle Type Ia supernovae, $H_0\!=\!73.24\! \pm\! 1.74\, {\rm km/s\, 
Mpc}$, which differs by 3$\sigma$ from the cosmic value $H_0\!=\!67.74\!\pm\! 
0.46\,{\rm km/s\ Mpc}$ obtained from the cosmic microwave background 
radiation by Planck assuming the standard $\Lambda$CDM cosmology. The 
measured superluminal motion of the jet also allows critical tests of the 
assumed production mechanism of SHBs in general and of SHB170817A in 
particular.
\end{abstract}

\pacs{98.70.Sa,97.60.Gb,98.20}

\maketitle

{\bf Introduction.} A $3\,\sigma$ level discrepancy exists between the 
measured local value of the Hubble constant provided by standard candle 
Type Ia supernovae$^1$, $H_0\!=\!73.24\!\pm\!1.74\,{\rm km/s\, Mpc}$,
and the cosmic value obtained from the  
Planck measurements$^2$ of the microwave background radiation assuming the 
standard $\Lambda$CDM cosmology$^2$, $H_0\!=\!67.74\! \pm\! 0.46\, {\rm km/s\, Mpc}$.
The origin of this discrepancy is still 
unknown. An independent method of measuring the local value of $H_0$ 
from gravitational wave (GW) sources accompanied by emission of 
electromagnetic waves (EMW) was  suggested before$^{3,4}$ 
and applied recently to obtain the value$^{5}$ 
$H_0\!=\!74^{+16}_{-8}\,{\rm km/s\, 
Mpc}$ from the GW emission from the nearby neutron stars merger (NSM) event$^6$ 
GW170817 followed by the short gamma ray burst$^7$ SHB170817A. Moreover,
in a recent paper$^8$ it was claimed that the VLBA and 
VLBI radio observations$^9$  of the mean apparent 
superluminal velocity of the jet in the short hard burst SHB170817A 
improves the measurement of the local Hubble constant 
based on the gravitational waves  emission from the NSM GW170817
event  and the electromagnetic localization of 
the counterpart short hard gamma ray burst SHB170817A$^{7,10}$. This 
claimed improvement, however, was  based$^8$ on a debatable model,
which  involved many free adjustable parameters and a jet direction  
which does not necessarily coincide with the axis of the orbit of the merging 
neutron stars$^8$. A coincidence between the axis  of the 
orbital motion and the direction of the emitted highly relativistic jet is 
observed in quasars, blazars and microquasars. 

Below, we present evidence from the combined GW170817/SHB170817A event
and the measured$^9$ apparent superluminal velocity of the 
the far off-axis jet launched in this event, which seems to support the 
local value$^1$ of the Hubble constant obtained from standard candle 
supernovae of type Ia, but is  different, roughly by $3\sigma$ from the 
cosmic value measured$^2$ by Planck, assuming the standard $\Lambda$CDM 
cosmology. We also show that the measured superluminal motion of the 
jet allows  critical tests of the assumed production mechanism of SHBs 
in general and of SHB170817A in particular.

{\bf Apparent Superluminal Motion Evidence.} 
The apparent  velocity in the plane of the sky 
of a highly 
relativistic compact (unresolved) source moving at a small redshift  
($1\!+\!z\!\approx\!1$) with a {\bf constant} bulk motion Lorentz 
factor $\gamma\!\gg\!1$ 
(i.e., $\beta\!=\!\sqrt{1-1/\gamma^2}\!\approx\!(1\!-\!1/2\gamma^2)\!
\approx\! 1$), 
and viewed from an angle $\theta$ relative 
to its direction of motion, is given by 
\begin{equation}
V_{app}\!=\! {\beta\, c\, sin\theta\over (1\!+\!z) (1\!-\!\beta\,cos\theta)}
\!\approx\! {c\, sin\theta\over (1\!-\!cos\theta)}\,,  
%\label(Eq1)
\end{equation}
which depends only on $\theta$.  In that case, the 
angular distance to  the afterglow source  satisfies  
\begin{equation} 
D_A\!\approx V_{app}\,\Delta t/\Delta\theta_s\!\approx\!c\, z/H_0
%\label{Eq2) 
\end{equation} 
where $\Delta\theta_s$ is the change in its angular location 
during the time $\Delta t$.
In the case of  SHB170817A in NGC 4993 at redshift$^{10}$  $z\!=\! 0.009783$, 
the angular  location of the source in the plane of the sky has 
changed during $\Delta t$=155 d (between day 75 and day 230) by 
$\Delta\theta_s\!=\!2.68\!\pm\!0.3$ mas$^9$.
Thus, assuming the local value$^1$ $H_0\!=\!73.24\!\pm\! 1.74\,{\rm km/s\, Mpc}$,
eq.(2) yields  $V_{app}\!\approx\! (4.0\!\pm\! 0.40)\,c$  and 
consequently  $\theta\approx 28\!\pm\! 2$ deg, which  follows from eq.(1).
This value of $\theta$ is in  agreement with the value $25\pm 4$ deg, which 
was obtained$^{11}$ from GW170817 and its electromagnetic location$^{10}$
assuming the local value$^1$ of $H_0$ obtained from Type Ia supernovae (SNeIa).
Eq.(2) yields  $V_{app}\!\approx\! (4.32\!\pm\! 0.43)\,c$,  and
consequently  $\theta\approx 26\!\pm\!2$ deg, which  follows from eq.(1)
for the Planck cosmic value$^2$ $H_0\!=\!67.74\!\pm\! 0.46\, {\rm km/s\, Mpc}$.
It is however in tension with $\theta\!=\!18\pm 4$ deg,  
which was obtained$^{11}$ from GW170817 and its electromagnetic location$^{10}$
assuming$^{11}$ the Planck cosmic value$^2$ of $H_0$ obtained from the MBR.

Note  that if $\gamma(t)$ is time dependent, then 
\begin{equation}
V_{app}\!=\!\beta\,\gamma\, \delta\, c\, sin\theta/(1\!+\!z) 
%\label(Eq3)
\end{equation}
where $\delta\!=\!1/\gamma\,(1\!-\beta\,\!cos\theta)$ is the Doppler factor 
of the source. Thus, $V_{app}(t)$  generally depends on both 
$\theta$  and $\gamma$ which are generally unknown.
But, as long as $\gamma^2\!\gg\! 1$,  $\theta^2\!\ll\!1$,
and  $\gamma^2\,\theta^2\gg\!1$,  $V_{app}$ remains constant in time 
and satisfies eq.(1). 

A constant superluminal velocity $V_{app}\!\approx\!c\, 
sin\theta/(1\!-\!cos\theta)$
was  predicted by the cannonball (CB) model of GRBs for far-off axis (low luminosity) 
GRBs  within a face-on spiral galaxies$^{12}$, 
where $\gamma^2\, \theta^2\!\gg\! 1$ is satisfied within the disk  
and remains so after escape into the galactic  halo.
Both  GRB980425$^{13}$ and SHB170817A$^{14}$  took place in such locations. 
The deceleration of a CB stops practically  when the CB moves out from the 
galactic disk into the low density galactic halo$^{12,15}$. 

Note in particular that the shape of the prompt emission and of
the afterglow, in the CB model, depends on the product $\gamma(0)\,\theta$
and not on the individual values of $\gamma(0)$ and $\theta$.  This
degeneracy has now been removed, for the first time, by
the measured$^9$ apparent superluminal motion of the jet in
SHB170817A, which yields $\theta\!=\!28$ deg. As shown below,
it indicates that the glory around NSMs
has its peak energy in the X-ray band. This is in contrast
to the eV peak energy of the optical glory in SN-GRBs,
{\bf   which, in the past, was wrongly assumed$^{15}$ also for SHBs}.

{\bf Falsibiable tests of the production mechanism of SHBs.}  In the CB 
model$^{16}$ the peak energy 
of the prompt emission produced by ICS of glory photons with a 
bremsstrahlung (or a cutoff power-law) spectrum,
$\epsilon\,(dn/d\epsilon)\propto exp(-\epsilon/\epsilon_p)$, is 
given by
\begin{equation}
(1\!+\!z)\, E_p\!=\,\gamma\, \delta \, \epsilon_p\, .
%\label(Eq4)
\end{equation}
In low luminosity SHBs, i.e., SHBs viewed far off-axis, $\gamma^2\theta^2\!\gg\!1$ 
and $\beta\!=\sqrt{1\!-\!1/\gamma^2}\!\approx\!1\!-\!1/2\gamma^2\!\approx\! 1$.
Thus, 
\begin{equation}
(1\!+\!z)\,E_p\!\approx\! \epsilon_p/(1-cos\theta). 
%label(Eq5)
\end{equation}
Hence, the observed$^{17}$ (T90) $E_p\approx 86\pm 19 $ keV 
and $\theta\approx 28$ deg yield an X-ray glory with
$\epsilon_p\!\approx\! 10.2 \pm 2.2$ keV. 

The viewing angle extracted from the observed 
superluminal motion of the jet in SHB170817A and the rise time 
$\Delta\!\approx 0.58$ s to peak value of its prompt emission 
pulse yield a glory radius 
\begin{equation}
R_g\!\approx\!{\gamma\, \delta\,c\, \Delta\over(1\!+\!z)}\!=\!
{c\,\Delta\over(1\!+\!z)(1\!-\!cos\theta)}\!\approx\!1.5\times 
10^{11} cm,
%label(Eq6)   
\end{equation}
Note also that the escape by diffusion of 
a  $10.2\pm 2.2$  keV glory/thermal light through the merger ejecta
surrounding the  newly born neutron star may explain 
the origin of the second prompt emission pulse in SHB170817A
with  $\sim 10.3\!\pm\!1.5$ keV grey body like spectrum$^{17}$.

Moreover, assuming that the glory and the CB Lorentz factor 
in SHBs are approximately standard candles, then the observed 
mean rise time $\langle\! \Delta\!\rangle\! \approx\! 25$ ms
of individual pulses in ordinary SHBs (where $\theta\!\approx\! 
1/\gamma$, i.e., $\delta\!\approx \!\gamma$) and eq.(6) yields  
\begin{equation}
\gamma^2\! \approx {R_g\,(1+z)\over \Delta\,c}\!\approx\!202\,,
%\label(Eq7)
\end{equation}   
i.e., $\gamma\!\approx\! 14.2$, both in SHB170817A and ordinary 
SHBs.  This value is roughly a factor 30 smaller than the mean value 
of $\gamma(0)$ of CBs launched in ordinary SN-GRBs$^{16}$. 

The isotropic equivalent energy of GRBs, in the CB model, 
satisfies$^{16}$ $E_{iso}\!\propto\! \epsilon_p\, \gamma\,\delta^3$.
Thus,  the CB model     
predicts that ordinary SHBs and SN-GRBs satisfy
\begin{equation}
\langle E_{iso}(SHB)\rangle \!\sim\!10^{-2}\,
\langle E_{iso}(SN\!-\!GRB)\rangle \!\sim\!10^{51}\,{\rm erg},
%\label{Eq8}
\end{equation}
and the far off-axis SHB170817A satisfies
\begin{equation}
{E_{iso}(170817A) \over \langle E_{iso}(SHB)\rangle}\!\approx\! 
{1\over \gamma^6\,(1\!-\!cos\theta)^3} \!\approx\!7.5\times 
10^{-5},
%\label(Eq9)   
\end{equation}
i.e., an $E_{iso}(170817A)\sim 7.5\times 10^{46}$ erg, 
compared to its observed value$^{17}$, $E_{iso}(170817A)\!
\approx\! (5.6\pm 1.1)\times 10^{46}$ erg. 

Note also that the peak energy of the far off-axis SHB170817A, 
as given by eq.(4), is related to the mean peak energy of ordinary SHBs  by
\begin{equation}
E_p(170817A)\approx \langle (1\!+\!z)\,E_p(SHB)\rangle/\gamma^2\,
(1-cos\theta) 
%\label(Eq10)
\end{equation}
where $\langle E_p(SHB)\rangle\approx  650$$^{17}$ keV. Assuming that 
the evolution function of SHBs is the same as that of long GRBs,
(which is the case if SHBs are produced by NSM of compact binaries 
in core fission during core collapse supernovae explosions 
of fast rotating massive stars$^{15}$), then $\langle 1\!+\!z\rangle\!\approx\!3$. 
Thus,  for $\gamma^2\!\approx\! 202$ as given by eq.(7), 
and $\theta\!=\!28$ deg, 
the right hand side of eq.(10) yields 82 keV, in agreement with 
the observed$^{17}$  $E_p\!=\!86\pm 19$ keV  ($T90=2.1$ s). 

{\bf Pulse Shape.} The observed  pulse-shape produced by ICS of glory 
photons with an exponentially cut off power law (CPL) spectrum, 
$dn_g/d\epsilon\!\propto\! \epsilon^{-\alpha}\,exp(-\epsilon/\epsilon_p)$ 
at redshift
$z$, by a CB is given approximately$^{15}$ by 
\begin{equation}
E{d^2N_\gamma\over 
dE\,dt}\!\propto\!{t^2\over(t^2\!+\!\Delta^2)^2} \, E^{1-\alpha}\,exp(-E/E_p(t)) 
\label{Eq11} 
\end{equation} 
where $\Delta$ is approximately the peak time of the pulse in the observer 
frame, which occurs when the CB becomes transparent to its internal
radiation, and $E_p\!\approx\! E_p(t\!=\!\Delta)$.
In eq.(11), the early temporal rise like $t^2$ is produced by its  increasing 
cross section, $\pi\, R_{CB}^2\!\propto\! t^2$, of the fast expanding CB when it 
is still opaque to radiation. When the CB 
becomes transparent to radiation due to its fast expansion, its 
effective cross section for ICS becomes constant.
That, and the density of the ambient photons, which for a distance  
$r\!=\!\gamma\,\delta\,c\,t/(1\!+\!z)\!>\! R$ decreases like 
$n_g(r)\!\approx\!n_g(0)\,(R/r)^2\!\propto\! t^{-2}$, produce the temporal 
decline like $t^{-2}$. If  CBs are  launched along the axis of a glory  
of torus-like pulsar wind nebula (PWN), or of an accretion disk with a 
radius $R$, then 
glory photons that intercept a CB at a distance $r$  from 
the center intercept it at a  lab angle $\theta_{int}$, which 
satisfies
$\cos\theta_{int} \!=\!-\! r/\sqrt{r^2\!+\! R^2}$. 
It yields a $t$-dependent peak energy,
\begin{equation}
E_p(t)\!=\!E_p(0)(1\!-\!t/\sqrt{t^2\!+\!\tau^2})\,,
\label{Eq12} 
\end{equation} 
with  $\tau\!=\!R\, (1+z)/\gamma\,\delta\,c\!\approx\!0.59$ s 
and $E_p\!\approx\! E_p(t\!\approx \!\Delta)$, where $\Delta$ 
is approximately 
the peak time of the pulse. For $\alpha$ not very different from 1,
integration of $d^2N(E,t)/dE\,dt$ from $E\!=\!E_m$ upwards yields
\begin{equation}
N(t,E\!>\!E_m)\!\propto\!{t^2\over (t^2+\Delta^2)^2}\,exp(\!-\!E_m/E_p(t)).
\label{Eq13}
\end{equation}
A best fit of eq.(13) to the observed pulse shape$^{17}$ for $E_m\!=\!50$ keV,
shown in Figure 1, returned  $\Delta\!=\!0.58$ s,
$\tau\!=\!0.65$ s, and $E_p(0)\!\approx\!320$ keV.
The best fit value of $\Delta$ yields 
$E_p\!\approx\! E_p(\Delta)\!=\!94$ keV compared to the observed$^{17}$  (T90)
$E_p\!=\!86\pm 19$ keV.
\begin{figure}[]
\centering
%\vspace{-2cm}
\epsfig{file=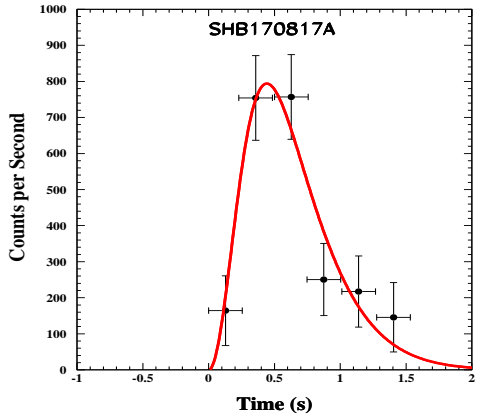,width=7.cm,height=6.cm}
\caption{Comparison of the pulse shape for $E_m\!=\!50$ keV
of the first pulse of SHB170817A$^{17}$ 
and the CB model pulse shape as given by eq.(13).}
%\label{fig1}
\end{figure}

{\bf Early-time Afterglow.}
The X-ray afterglow of ordinary SHBs are well
explained by PWN emission  powered  by the rotational energy loss
through magnetic dipole radiation (MDR), relativistic winds and high 
energy particles of the newly born millisecond pulsars (MSPs) in NSMs
In a steady state, the X-ray  luminosity  powered by 
the spin-down of  the MSP has the universal form$^{18}$ 
\begin{equation}
L(t)\!=\!L(0)\!1/(1\!+\!t/t_b)^2, 
%\label{Eq14}
\end{equation}
where  $t_b\!=\!P/2\, \dot{P}$
and $P$ being the period of the newly born pulsar at birth.
A best fit of eq.(14) to the bolometric light curve$^{19}$ of SHB170817A 
shown in Figure 2 yields
$L(0)\!=\!2.27\times 10^{42}$ erg/s, $t_b\!=\!1.15$ d, with an entirely 
satisfactory $\chi^2/dof\!=\!1.04$.

\begin{figure}[]
\centering
\epsfig{file=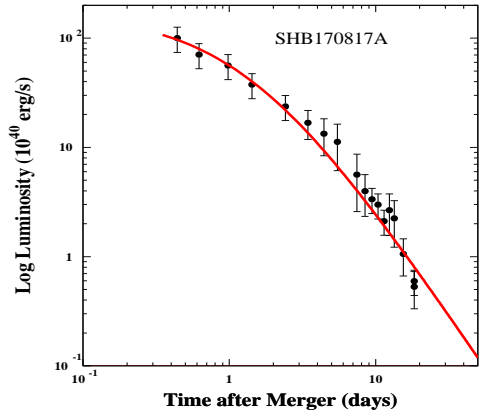,width=7.cm,height=6.cm}
\caption{Comparison between the observed$^{19}$ bolometric lightcurve
of SHB170817A and the best fit CB model lightcurve assuming MSP 
remnanant, as give by eq.(14).}
%\label{Fig2}
\end{figure}

{\bf The Synchrotron Afterglow.}  In the CB model,
the electrons that enter the CB with a Lorentz factor
$\gamma(t)$ in its rest frame are Fermi accelerated there, and
cool by emission of synchrotron radiation - an isotropic
afterglow in the CBs rest frame. As for the rest of the
CBs radiations, the emitted photons are beamed into a
narrow cone along the CBs direction of motion, their
arrival times are aberrated, and their energies boosted
by the Doppler factor $\delta(t)$ and redshifted by the cosmic
expansion. 
The observed spectral energy density of the unabsorbed
synchrotron afterglow produced by a CB has the form
(e.g., eq.(1) in Ref. [20]),
\begin{equation}
F_\nu\!\propto\! n^{(1\!+\!\beta_\nu)/2)}\, [\gamma(t)]^{3\beta_\nu\!-\!1}
[\delta(t)]^{\beta_\nu\!+\!3}\nu^{-\beta_\nu}\,,
%\label{Eq15}
\end{equation} 
where $n(t)$ is the density of the medium encountered by
the CB at time $t$ and $\beta_\nu$ is the spectral index of the 
emitted radiation at a frequency $\nu$.  For a constant density,
the deceleration of the CB yields a late time $\gamma(t)\propto 
t^{-1/4}$ and $\delta(t)\!=\! 1/\gamma(t)\,(1\!-\!cos\theta)\!
\propto\! t^{1/4}/(1-cos\theta)$ .  
Consequently, the apparent superluminal velocity of a CB in
a constant low-density environment, stays constant, as
long as $\beta\!\approx\!1$, while its late-time afterglow 
luminosity increases like $t^{(1-\beta_\nu/2)}$. When the CB exits 
the disk into the halo, it turns into a fast decay  
$\propto\![n(r)]^{(1\!+\!\beta_\nu)/2}$. Approximating the disk density
perpendicular to the disk by $n(z)\!=\!n(0)/(1\!+\!exp(z/d))$
where $d$ is the surface depth of the disk, the lightcurve of 
the  afterglow of  SHB170817A can be approximated by, 
\begin{figure}[]
\centering
%\vspace{-2cm}
\epsfig{file=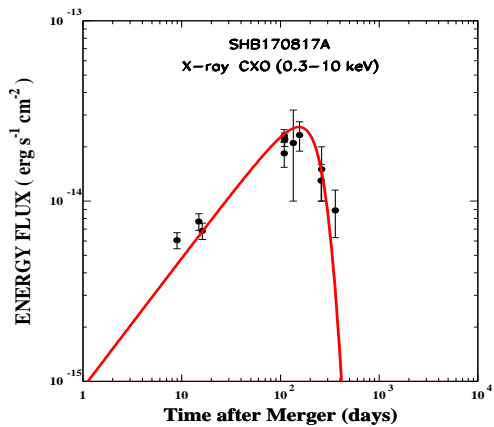,width=7.cm,height=6.cm}
\caption{The lightcurve of the X-ray 
afterglow of SHB170817A  measured$^{22}$ with the CXO 
and the lightcurve predicted by eq.(16)  for $\beta_X=0.56$,
$t_e\!=\!245.6$ d  and $w=63.4$ d.}
\label{fig3}
\end{figure}
\begin{equation}
F_\nu(t)\!\propto\!{(t/t_e)^{1\!-\!\beta_\nu/2}\, \nu^{-\beta_\nu}\over
[1+exp[(t\!-\!t_e)/w]]^{(1\!+\!\beta_\nu)/2}} 
%\label{Eq16}
\end{equation}
where $t_e$ is roughly the escape time of the CB from the 
galactic disk into the halo after its launch, 
and $w$ is the crossing time of the surface of the disk. 
Such a behavior of the late-time afterglow may 
have been observed$^{21~}$ in GRB98042515. It is compared in 
Figures 3,4 to the observed late-time X-ray$^{22}$ and 
radio$^{23}$ afterglows of SHB170817A.

\begin{figure}[]
\centering
%\vspace{-2cm}
\epsfig{file=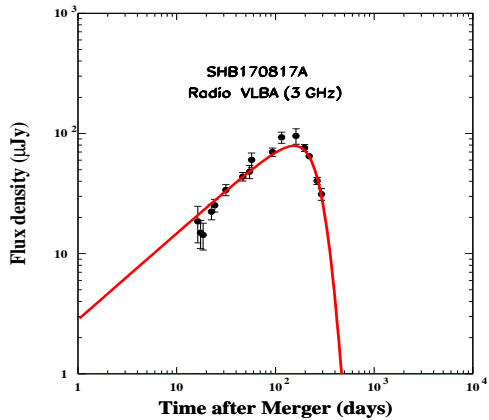,width=7.cm,height=6.cm}
\caption{Left: The measured$^{23}$
lightcurve of the 3 GHz radio afterglow of SHB170817A   
and the lightcurve predicted by the CB mode $\beta_r=0.56$,
$t_e\!=\!245.6$ d  and $w=63.4$ d.}                         
\label{fig4}
\end{figure}

{\bf Conclusions.}
The first successful measurement of the apparent superluminal motion of the 
highly relativistic jets, which produce the prompt emission and the beamed 
afterglow of GRBs and SHBs, finally took place$^9$ two decades after the 
discovery of the afterglow of GRBs at lower frequencies$^{24}$. This recent 
achievement not only is an important contribution towards resolving the long 
standing debate of what is the production mechanism of SHBs and their 
afterglows, it also appears to confirm a fundamental physical difference 
between the local and the cosmic values of the Hubble constant. Few more 
Ligo virgo detections of nearby neutron stars merger followed by far-off axis 
SHBs with accurate VLBA and VLBI measurement of the superluminal motion of 
their jets will be required to confirm that beyond doubt.

\end{document}